\documentclass{article}
\usepackage{xurl} 
\usepackage{spconf,amsmath,graphicx,hyperref}
\usepackage[acronym]{glossaries-extra}
\usepackage{subcaption} 
\usepackage{booktabs} 
\usepackage[table]{xcolor} 
\usepackage{pgf} 
\usepackage{tipa} 

\title{Beyond Utterance-Level Speaker Similarity: Perception-Aligned Segmental Evaluation of Accent in Speech Generation}
\name{Jinzuomu Zhong, Alice Ross, Catherine Lai, Korin Richmond}
\address{Centre for Speech Technology Research, University of Edinburgh}

\makeglossaries
\setabbreviationstyle[acronym]{long-short}
\setabbreviationstyle[onlyshort]{short}

\newacronym{zstts}{ZS-TTS}{Zero-Shot Text-to-Speech}
\newacronym{sota}{SOTA}{State-of-the-art}
\newacronym[category=onlyshort]{tts}{TTS}{Text-to-Speech}

\newacronym{mos}{MOS}{Mean Opinion Score}
\newacronym{smos}{SMOS}{Similarity MOS}

\newacronym{aid}{AID}{Accent Identification}
\newacronym{sv}{SV}{Speaker Verification}
\newacronym{ppgs}{PPGs}{Phonetic Posteriorgrams}
\newacronym{dtw}{DTW}{Dynamic Time Wrapping}
\newacronym{rmse}{RMSE}{Root Mean Squared Error}
\newacronym{js}{JS}{Jensen–Shannon}

\newacronym{mfa}{MFA}{Montreal Forced Aligner}
\newacronym{g2p}{G2P}{Grapheme-to-Phoneme}

\newacronym{srcc}{SRCC}{Spearman's Rank Correlation Coefficient}

\begin{document}
\ninept
\maketitle
\begin{abstract}
Utterance-level similarity ratings do not reveal where perceived accent differences occur in generated speech. We introduce a segmental evaluation protocol combining speaker and accent similarity ratings with subword annotations of perceived accent differences provided by listeners from target accent regions. Using newly collected recordings, 47 retained listeners evaluate 200 content-matched reference--candidate pairs spanning four Scottish and Australian voices, two zero-shot TTS systems, and an articulation-based voice conversion system. Subword annotations locate salient pronunciation differences and sometimes yield different system rankings from similarity ratings. Speaker and accent ratings remain strongly associated ($\rho = 0.71$), highlighting the value of examining local perceptual evidence beyond these scores. Evaluation of seven objective distances reveals weak correspondence with subword selections, with correlations reaching at most $\rho = 0.22$. We release code, data, and an interactive demo to support the development of perceptually grounded segmental accent evaluation.
\end{abstract}

\vspace{-0.5em}
\section{Introduction}
\label{sec:intro}
\vspace{-0.5em}

\gls{zstts} systems have made substantial progress in voice cloning, with recent models reporting utterance-level speaker similarity scores comparable to, or even exceeding, those of ground-truth recordings \cite{wang2025maskgct, chen2025f5tts}. However, claims of ``human-parity'' based on these scores do not necessarily reflect how speakers perceive their own cloned voices in personalised applications, such as assistive communication for users with voice impairment. Recent perception studies suggest that, when evaluating unfamiliar speakers' voices on common benchmark datasets, listeners may find speech generated by \gls{sota} \gls{tts} systems perceptually indistinguishable from human recordings \cite{srinivasavaradhan2025the}. Yet, in an own-voice evaluation setup, speakers can distinguish real recordings from synthetic counterparts with 87\% accuracy, utilising accent and prosody as main cues \cite{sanchez2026when}. This contrast raises a crucial question: how can we design listening tests and objective metrics that better reflect users' perceptions of cloned voices?

Accent\footnote{Here, we define \textit{accent} as a pattern of pronunciation characteristics used by an individual speaker or, more generally, by a community or social group. It encompasses both segmental features such as vowel and consonant realisations and suprasegmental features such as intonation and rhythm \cite[Ch.~1]{wells1982accents}.} provides a compelling case through which to investigate this discrepancy. It conveys regional, social, and linguistic identity \cite{michel2025s}, yet speech-generation systems often produce an accent that deviates from the reference recording, a phenomenon termed accent hallucination \cite{zhong2025accentbox}. Growing efforts towards high-fidelity accent generation \cite{zhong2025accentbox,li2025scalable,wang2026joycent} make reliable evaluation increasingly important.

Recent work has explored complementary approaches to accent evaluation. CodecMOS-Accent \cite{huang2026codecmosaccent} uses accent \gls{smos}, alongside other metrics, to benchmark speech codec resynthesis and \gls{tts} systems. However, its ratings are collected predominantly from American listeners, whose familiarity with the target accents is not established. It therefore remains unclear how well these ratings reflect pronunciation differences perceived by speakers from the target accent communities. Meanwhile, the enhanced XAB listening test in \cite{zhong2025pariwise} supplements pairwise judgements with character-level annotations of perceived accent differences, aiming to improve judgement consistency. These annotations provide a means of locating segmental accent differences, but are not analysed further. Pairwise comparisons also become increasingly costly as the number of systems grows, and the study covers limited systems and listeners.


To address these limitations, we combine utterance-level speaker and accent \gls{smos} with subword annotations of perceived accent differences. Building on \cite{zhong2025pariwise}, we analyse segment annotations as an evaluation outcome rather than as an auxiliary task, and examine their correspondence with utterance-level ratings and objective metrics. Our contributions are threefold:

\noindent 1) We introduce a subword-level annotation protocol that captures listeners' perceptions of segmental accent differences, complementing existing utterance-level speaker and accent \gls{smos} ratings.
\vspace{0.25em}

\noindent 2) We construct a perceptually annotated evaluation set using newly collected, content-matched reference--candidate pairs and listeners from the target accent regions, and analyse how segmental annotations relate to utterance-level ratings and system rankings.
\vspace{0.25em}

\noindent 3) We assess seven objective distances against utterance-level ratings and subword-level annotations, using temporal alignment and frame-to-subword aggregation for local comparison. Our analysis identifies limited correlation between the evaluated distance metrics and listeners' segmental annotations.
\vspace{0.25em}

To facilitate reproducibility and future investigation, we release the code, data, and an interactive results demo here\footnote{
    Code: \url{https://github.com/jzmzhong/segmental_accent_evaluation}\\
    Data: \url{https://huggingface.co/datasets/jzmzhong/segmental_accent_annotation}\\
    Demo: \url{https://jzmzhong.github.io/Segmental-Accent-Evaluation/results.html}
}.

\vspace{-0.5em}
\section{Grounding Human Perception: Subjective Rating and Segmental Annotation}
\label{sec:method_subjective}
\vspace{-0.5em}

\subsection{Three-Stage: Recording, Synthesis, and Evaluation}
\label{ssec:subjective_pipeline}
\vspace{-0.5em}

To ground accent evaluation in human perception, we conduct a three-stage data collection: recording natural speech, synthesising speech, and collecting listener ratings and segmental annotations.

In the first stage, participants record prepared utterances and provide demographic information. Collecting new recordings allows us to sample contemporary regional varieties of English and document participants’ linguistic and residential backgrounds.
This provides greater control over speaker recruitment and recording recency than relying on existing benchmark corpora.
Newly collected recordings also reduce the risk of overlap with the training data of the speech generation systems under evaluation.

In the second stage, we apply DeepFilterNet \cite{schroeter2022deepfilternet2} to reduce background noise.
We then generate candidate speech using the TTS systems under evaluation, each conditioned on a randomly selected recording from the same speaker as voice prompt. Unlike existing benchmarks such as CoedcMOS-Accent \cite{huang2026codecmosaccent} and Seed-TTS-Eval\footnote{\url{https://github.com/BytedanceSpeech/seed-tts-eval}}, each generated utterance is paired with a \textbf{content-matched reference}, rather than the voice prompt. Matching the linguistic content facilitates direct comparison of corresponding segments for accent similarity ratings and accent clue annotation.

\begin{figure}[h!]
    \centering
    \includegraphics[width=0.7\linewidth]{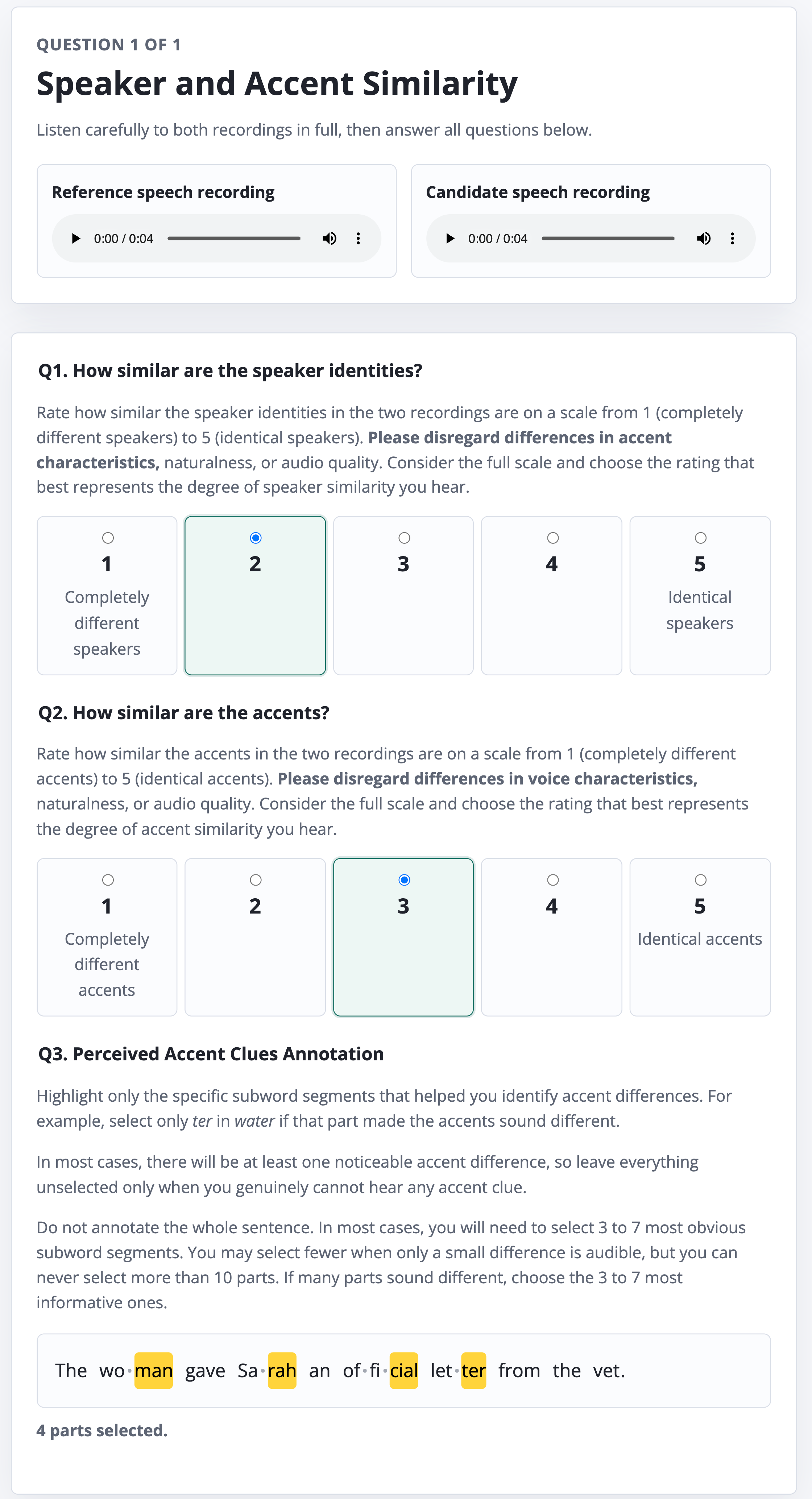}
    \captionsetup{justification=centering}
    \caption{Listening-test interface for speaker \gls{smos}, accent \gls{smos}, and segmental annotation of perceived accent differences.}
    \vspace{-2em}
    \label{fig:annotation_interface}
\end{figure}

In the third stage, listeners from the \textbf{same accent region} as the target speaker separately rate speaker and accent \gls{smos} and highlight transcript segments that reveal perceived accent differences. 
Fig.~\ref{fig:annotation_interface} shows the evaluation interface.

\vspace{-0.5em}
\subsection{Iterative Listening Test Design}
\label{ssec:subjective_design}
\vspace{-0.5em}

We refine the instructions and annotation procedures iteratively through pilot studies, focusing on separate assessment of speaker identity and accent, and consistent selection of accent clues.

\vspace{0.25em}
\noindent\textbf{Disentangling accent and speaker similarity ratings:} 
Listeners are explicitly instructed to judge speaker identity independently of accent and accent independently of voice characteristics (Q1 and Q2 in Fig.~\ref{fig:annotation_interface}). Alongside the evaluated TTS systems, we include speech reconstructed using SPARC \cite{cho2024coding}, an articulation-based voice conversion system. SPARC reconstructs the content-matched reference using a speaker embedding extracted from another utterance by the same speaker. Its articulation-based representation and lightweight speaker encoding motivate its use as an accent-preserving reference with reduced speaker fidelity, providing an expected top-line for accent similarity and bottom-line for speaker similarity.

\vspace{0.25em}
\noindent\textbf{Encouraging consistent segmental annotation:} Motivated by the low inter-annotator agreement observed for character-level annotations in \cite{zhong2025pariwise}, we use subword segments as a coarser annotation unit. Subword segments approximate syllables in the written transcript without requiring annotators to have prior phonetic training. We use Pyphen \footnote{\url{https://github.com/Kozea/Pyphen}} for subword segmentation, with subsequent manual review. Pilot studies reveal substantial variation in annotation tendencies: some listeners select very few segments, whereas others mark large portions of the transcript. We therefore encourage listeners to select 3--7 informative segments in most cases, with a maximum of 10 segments allowed (Q3 in Fig.~\ref{fig:annotation_interface}). Listeners may leave the transcript unmarked when they perceive no accent difference.

\vspace{-0.5em}
\section{Finding Instrumental Correlates: Adapting Objective Metrics for Segmental Comparison}
\label{sec:method_objective}
\vspace{-0.5em}

To investigate whether objective metrics correlate with listeners' perceptions, we first compute utterance-level distances using seven metrics (Sec.~\ref{ssec:method_objective_utterance}). We then derive frame-level distances (Sec.~\ref{ssec:method_objective_frame}) and map them to the subword segments annotated by listeners (Sec.~\ref{ssec:method_objective_frame2subword}), enabling correlation analysis at both utterance and subword levels.

\vspace{-0.5em}
\subsection{Utterance-Level Distances}
\label{ssec:method_objective_utterance}
\vspace{-0.5em}

\noindent\textbf{Speaker and accent representations:}
We use two \gls{sv} models, ECAPA-TDNN \cite{desplanques2020ecapa} and WavLM-SV \cite{chen2022wavlm}, for speaker embeddings, and two \gls{aid} models, GenAID \cite{zhong2025accentbox} and Voxlect \cite{feng2026voxlect} for accent embeddings.
For each model, we obtain reference and candidate utterance embeddings in the last latent layer and calculate their cosine distances.

\vspace{0.25em}
\noindent\textbf{Phonetic and acoustic features:}
We use three metrics to examine pronunciation and prosodic differences: PPG distance, formant distance, and F0 distance. We extract \gls{ppgs} using a model trained on General American English \cite{churchwell2024ppgs}, which predicts a distribution over phonetic categories at each frame. We also extract the first three formants (F1--F3) to characterise differences in vowel realisation \cite{ghorshi2006comparative}, and fundamental frequency (F0) to characterise prosodic differences.

To account for differences in length, we align the feature sequences of each reference--candidate pair using \gls{dtw}. For PPGs, we use \gls{js} distance as both the alignment cost and the local distance between aligned frames. For acoustic features, we compute a shared alignment using the \gls{rmse} between frame-level vectors $[\mathrm{F0},\mathrm{F1},\mathrm{F2},\mathrm{F3}]$. Along this shared path, we calculate separate local distances using absolute error for F0 and \gls{rmse} across F1--F3 for formants.
For each of the three metrics, we obtain the utterance-level distance $D_{\mathrm{utt}}$ by averaging the distances over all aligned frame pairs:
\vspace{-0.5em}
\begin{equation}
    D_{\mathrm{utt}} =
    \frac{1}{|\mathcal{P}|}
    \sum_{(i,j)\in\mathcal{P}}
    \delta(\mathbf{x}_i,\mathbf{y}_j),
\vspace{-0.5em}
\end{equation}
where $\mathcal{P}$ is the DTW alignment path, $\mathbf{x}_i$ and $\mathbf{y}_j$ are the reference and candidate frame features, and $\delta$ is the metric-specific local distance: \gls{js} distance, absolute F0 error, or \gls{rmse} across F1--F3.

\vspace{-0.5em}
\subsection{Frame-Level Distances}
\label{ssec:method_objective_frame}
\vspace{-0.5em}

For speaker and accent embeddings, utterance-level distances compare embeddings obtained after temporal pooling. To calculate frame-level distances, we instead extract the representations immediately before pooling and align the reference and candidate sequences using \gls{dtw}, with cosine distance as both the alignment cost and the local distance. Although these representations are not explicitly trained to measure local speaker or accent similarity, they preserve a temporal sequence that may support segmental comparison. For PPGs, F0, and formants, we retain the alignment paths and local distance functions used for utterance-level analysis.

For all seven metrics, we assign distance $d_j$ to each candidate frame by averaging the local distances over its aligned reference frame(s):
\vspace{-1em}
\begin{equation}
    d_j =
    \frac{1}{|\mathcal{A}_j|}
    \sum_{i\in\mathcal{A}_j}
    \delta(\mathbf{x}_i,\mathbf{y}_j),
    \qquad
    \mathcal{A}_j = \{i : (i,j)\in\mathcal{P}\},
\vspace{-1em}
\end{equation}
where $\mathcal{A}_j$ contains a set of reference frame indices $i$ aligned to candidate frame $j$. This produces one distance value per candidate frame for each metric.

\vspace{-0.5em}
\subsection{Mapping Frame-Level Distances to Subword Annotations}
\label{ssec:method_objective_frame2subword}
\vspace{-0.5em}

To compare \textit{frame}-level objective distances with \textit{subword}-level subjective annotations, we map acoustic frames to the subword segments used in the perceptual task. We first obtain the phone boundaries of each utterance using \gls{mfa} \cite{mcauliffe2017mfa}, aligning \textit{frames} to \textit{phones}. We then use a \gls{g2p} aligner \cite{richmond2009lts}, a probabilistic alignment algorithm that scores candidate letter-to-phone(s) correspondences against a pre-computed table of empirical alignment probabilities, to align \textit{phones} to letters of each word, and consequently onto the \textit{subword} segments used for annotation. Both alignment stages use an accent-independent pronunciation lexicon (with ARPABET phones), avoiding the reliance on accent-specific lexicon for each accent.

For each metric, we derive three alternative distances from the candidate frames associated with each subword segment: the maximum frame-level distance (\textbf{Max}), the mean frame-level distance (\textbf{Mean}), and the distance at the frame nearest the segment's temporal midpoint (\textbf{Midpoint}). These quantify the largest discrepancy within the segment, its average discrepancy, and the discrepancy at its centre, respectively. We compare these subword-level distances with the corresponding listener annotations.

\vspace{-0.5em}
\section{Experiments}
\label{sec:experiments}
\vspace{-0.5em}

\subsection{Subjective Listening Tests}
\label{ssec:exp_subjective}
\vspace{-0.5em}

\textbf{Test stimuli and systems:} For recording, participants read two passages commonly used in phonetic research on English varieties: \textit{Comma Gets a Cure}\footnote{\url{https://www.dialectsarchive.com/comma-gets-a-cure}} and \textit{The Boy Who Cried Wolf} \cite{deterding2006north}. For synthesis, we select two representative \gls{sota} \gls{zstts} systems: MaskGCT \cite{wang2025maskgct} and F5-TTS \cite{chen2025f5tts}, representing two-stage and one-stage acoustic modelling, respectively. Both systems achieve top speaker similarity performances on the TTSDS2 benchmark \cite{minixhofer2026ttsds2}.

For evaluation, to limit listener fatigue, we design the test to last approximately 40 minutes and use the first 20 utterances of \textit{Comma Gets a Cure}. We synthesise these utterances using a randomly selected recording from \textit{The Boy Who Cried Wolf} by the same speaker as the voice prompt. We additionally reconstruct the first 10 utterances using SPARC as a reference system. Each listener evaluates 50 content-matched reference--candidate pairs: 20 from MaskGCT, 20 from F5-TTS, and 10 from SPARC, comprising 1,035 subwords.

\vspace{0.25em}
\noindent\textbf{Participants recruitment and filtering:} We evaluate four target voices: one male and one female speaker from each of Scotland and Australia. For each voice, we recruit 15 listeners through Prolific from the corresponding accent region, each of whom rates and annotates all 50 pairs, yielding 60 listeners before exclusions.

We apply two listener exclusion criteria motivated by the pilot observations in Sec.~\ref{ssec:subjective_design}. First, we exclude listeners whose mean number of selected segments across all evaluation trials is below 2 or above 8, , identifying unusually sparse or extensive annotation relative to the requested selection range. Second, we exclude listeners who assign identical speaker and accent similarity ratings on more than 80\% of trials, flagging possible conflation of the two perceptual dimensions. After exclusions, 47 listeners remain, contributing 4,700 \gls{smos} ratings and approximately 48,600 binary subword annotation decisions across 200 reference--candidate pairs.

\vspace{-0.5em}
\subsection{Subjective-Objective Correlation}
\label{ssec:exp_correlation}
\vspace{-0.5em}

\noindent\textbf{Processing subjective scores:}
We calculate all subjective scores using retained listeners' responses. For each reference--candidate pair, we average speaker and accent similarity ratings across listeners to obtain S-SPK-SMOS and S-ACC-SMOS, respectively. We calculate S-Anno-Freq as the mean percentage of subword segments selected by each listener for that pair. At the subword level, we calculate S-Anno-Score -- the proportion of listeners who selected the segment, with higher values indicating greater agreement that the segment reveals an accent difference.

\vspace{0.25em}
\noindent\textbf{Calculating correlation:}
At the utterance level, we calculate pairwise \gls{srcc} among the three subjective scores to examine the relationships between speaker similarity, accent similarity, and annotation frequency. We then calculate \gls{srcc} between each of the seven objective distances in Sec.~\ref{ssec:method_objective_utterance} and each subjective score across all 200 reference--candidate pairs.
At the subword level, we evaluate the three aggregation methods defined in Sec.~\ref{ssec:method_objective_frame2subword}: \textbf{Mean}, \textbf{Max}, and \textbf{Midpoint}. For each of the seven metrics and each scoring method, we calculate \gls{srcc} between the resulting subword distances and S-ACC-Anno across 4,140 subword instances in the 200 pairs. Both analyses pool observations across the four target voices and three speech generation systems.

\begin{table*}[t!]
\centering
\small
\setlength{\tabcolsep}{4pt}
\renewcommand{\arraystretch}{1.1}
\resizebox{\textwidth}{!}{
    \begin{tabular}{@{}l*{11}{c}@{}}
    \toprule
    & \multicolumn{2}{c}{Inter-annotator agreement $\uparrow$}
    & \multicolumn{3}{c}{S-Anno-Freq (\%) $\downarrow$}
    & \multicolumn{3}{c}{S-ACC-SMOS $\uparrow$}
    & \multicolumn{3}{c}{S-SPK-SMOS $\uparrow$} \\
    \cmidrule(lr){2-3}
    \cmidrule(lr){4-6}
    \cmidrule(lr){7-9}
    \cmidrule(l){10-12}
    Speaker
    & Fleiss' $\kappa$ & Krippendorff's $\alpha^*$
    & F5-TTS & MaskGCT & SPARC
    & F5-TTS & MaskGCT & SPARC
    & F5-TTS & MaskGCT & SPARC \\
    \midrule
    
    Australian female
    & 0.184 & 0.192
    & 17.95\,{\scriptsize$\pm$\,1.45}
    & 20.11\,{\scriptsize$\pm$\,1.88}
    & \textbf{15.92}\,{\scriptsize$\pm$\,1.98}
    & \textbf{3.14}\,{\scriptsize$\pm$\,0.16}
    & 2.94\,{\scriptsize$\pm$\,0.17}
    & 2.77\,{\scriptsize$\pm$\,0.24}
    & \textbf{3.62}\,{\scriptsize$\pm$\,0.16}
    & 3.30\,{\scriptsize$\pm$\,0.17}
    & 1.14\,{\scriptsize$\pm$\,0.11} \\
    
    Australian male
    & 0.234 & 0.240
    & 22.07\,{\scriptsize$\pm$\,1.53}
    & 22.37\,{\scriptsize$\pm$\,1.44}
    & \textbf{19.95}\,{\scriptsize$\pm$\,2.15}
    & 2.81\,{\scriptsize$\pm$\,0.17}
    & \textbf{3.16}\,{\scriptsize$\pm$\,0.17}
    & 3.10\,{\scriptsize$\pm$\,0.27}
    & 3.11\,{\scriptsize$\pm$\,0.18}
    & \textbf{3.51}\,{\scriptsize$\pm$\,0.17}
    & 2.99\,{\scriptsize$\pm$\,0.27} \\
    
    Scottish female
    & 0.232 & 0.237
    & 23.92\,{\scriptsize$\pm$\,1.69}
    & 23.57\,{\scriptsize$\pm$\,1.55}
    & \textbf{17.05}\,{\scriptsize$\pm$\,2.23}
    & 1.98\,{\scriptsize$\pm$\,0.13}
    & 2.54\,{\scriptsize$\pm$\,0.15}
    & \textbf{3.91}\,{\scriptsize$\pm$\,0.18}
    & 2.32\,{\scriptsize$\pm$\,0.14}
    & 2.65\,{\scriptsize$\pm$\,0.15}
    & \textbf{3.73}\,{\scriptsize$\pm$\,0.19} \\
    
    Scottish male
    & 0.184 & 0.187
    & 22.74\,{\scriptsize$\pm$\,1.31}
    & 22.83\,{\scriptsize$\pm$\,1.36}
    & \textbf{18.20}\,{\scriptsize$\pm$\,2.07}
    & 2.06\,{\scriptsize$\pm$\,0.13}
    & 2.48\,{\scriptsize$\pm$\,0.13}
    & \textbf{3.57}\,{\scriptsize$\pm$\,0.18}
    & 2.22\,{\scriptsize$\pm$\,0.14}
    & 2.41\,{\scriptsize$\pm$\,0.15}
    & \textbf{2.83}\,{\scriptsize$\pm$\,0.21} \\
    \bottomrule
    \end{tabular}
}
\vspace{-0.5em}
\captionsetup{justification=centering}
\caption{Inter-annotator agreement and subjective evaluation results by target speaker. Best system for each target speaker within each metric is \textbf{bolded}. $*$: Krippendorff's $\alpha$ calculation excludes utterances with no or all subwords annotated.}
\label{tab:subjective_results}
\vspace{-2em}
\end{table*}

\vspace{-0.5em}
\section{Results \& Analysis}
\label{sec:results}
\vspace{-0.5em}

\subsection{Overall Results}
\vspace{-0.5em}

Table~\ref{tab:subjective_results} summarises inter-annotator agreement and subjective evaluation results by target speaker and synthesis system.
Annotators show modest agreement on the segmental annotation task, with Fleiss' $\kappa$ ranging from 0.184 to 0.234 across speakers. After excluding responses marking no or all subwords, Krippendorff's $\alpha$ ranges from 0.187 to 0.240. These values are numerically comparable to agreement reported for related perceptual annotation tasks: prosodic error annotation ($\alpha=0.18$--$0.28$ \cite{gutierrez2021location}) or prominence annotation ($\kappa=0.18$--$0.24$ \cite{bishop2020phonology}).

Both accent and speaker \gls{smos} favour F5-TTS for the Australian female speaker, MaskGCT for the Australian male speaker, and SPARC for both Scottish speakers, with identical system rankings for three of the four target speakers. In contrast, SPARC has the lowest proportion of annotated subwords for all four speakers, indicating fewer marked segmental accent differences and aligning with its intended role as an accent-preserving reference. The clearest divergence occurs for the Australian female speaker: SPARC receives the lowest accent and speaker SMOS (2.77 and 1.14, respectively), yet also has the lowest annotation rate (15.92\%). The similar rankings produced by the two SMOS measures are consistent with overlap between accent and speaker similarity judgements despite instructions to assess them separately. Segmental annotation provides a complementary perspective by directing listeners towards specific pronunciation differences.

\vspace{-1em}
\subsection{Utterance-Level Correlation}
\vspace{-1em}
\begin{table}[h!]
    \centering
    \setlength{\tabcolsep}{4pt}
    \renewcommand{\arraystretch}{1.1}
    \resizebox{\columnwidth}{!}{
        \begin{tabular}{lccc}
        \toprule
        Utterance-level Metric & S-ACC-Freq$^*$ & S-ACC-SMOS & S-SPK-SMOS \\
        \midrule
        S-ACC-Freq$^*$ & \cellcolor{red!100}1.00 &  &  \\
        S-ACC-SMOS & \cellcolor{red!48}0.48 & \cellcolor{red!100}1.00 &  \\
        S-SPK-SMOS & \cellcolor{red!27}0.27 & \cellcolor{red!71}0.71 & \cellcolor{red!100}1.00 \\
        \midrule
        O-F0$^*$ & \cellcolor{blue!7}-0.07 & \cellcolor{red!12}0.12 & \cellcolor{red!5}0.05 \\
        O-Formants$^*$ & \cellcolor{blue!9}-0.09 & \cellcolor{red!30}0.30 & \cellcolor{red!27}0.27 \\
        O-PPG$^*$  & \cellcolor{red!22}0.22 & \cellcolor{red!15}0.15 & \cellcolor{blue!25}-0.25 \\
        O-SPK (ECAPA-TDNN)$^*$ & \cellcolor{red!18}0.18 & \cellcolor{blue!32}-0.32 & \cellcolor{blue!11}-0.11 \\ 
        O-SPK (WavLM-SV)$^*$ & \cellcolor{red!30}0.30 & \cellcolor{blue!9}-0.09 & \cellcolor{red!5}0.05 \\ 
        O-ACC (GenAID)$^*$ & \cellcolor{red!9}0.09 & \cellcolor{red!11}0.11 & \cellcolor{red!13}0.13 \\ 
        O-ACC (Voxlect)$^*$ & \cellcolor{red!46}0.46 & \cellcolor{red!27}0.27 & \cellcolor{red!8}0.08 \\ 
        \bottomrule
        \end{tabular}
    }
    \vspace{-0.5em}
    \captionsetup{justification=centering}    
    \caption{\textit{Utterance}-level \gls{srcc} between evaluation metrics across 200 reference--candidate pairs. Distance metrics (marked with $*$) are negated to align with similarity scores. Larger positive coefficients indicate stronger ranking agreement.}
    \label{tab:corr_utt}
\end{table}
\vspace{-1em}
\noindent Utterance-level correlations are shown in Tab.~\ref{tab:corr_utt}. Among the subjective metrics, S-ACC-SMOS and S-SPK-SMOS are strongly correlated ($\rho=0.71$), consistent with the close association between accent and speaker \gls{smos} reported in \cite{huang2026codecmosaccent}. To investigate whether the high correlation is caused by actual covariation between accent and speaker similarity in the evaluated stimuli, or by the entangled perceptual judgements by listeners, we separately analyse 40 SPARC reference–candidate pairs. Informal expert listening confirms that these candidates largely preserve the reference accent and articulation while exhibiting noticeable artefacts in timbre. Within this subset, S-ACC-SMOS and S-SPK-SMOS remain quite strongly correlated ($\rho=0.57$), suggesting persistent overlap between accent and speaker similarity judgements despite the iterative refinements intended to distinguish these perceptual dimensions in Sec.~\ref{ssec:subjective_design}.

Negated S-ACC-Freq correlates moderately with S-ACC-SMOS ($\rho=0.48$) and more weakly with S-SPK-SMOS ($\rho=0.27$), suggesting closer alignment with perceived accent similarity. The partial agreement with the two \gls{smos} ratings indicates that segmental
annotation may provide complementary perceptual information not fully
reflected in holistic ratings. Qualitative inspection supports the accent relevance of the annotations: e.g.\ most listeners successfully identify pronunciation differences such as
\textipa{["fju:taIl]} in the reference versus \textipa{["fju:t@l]} in the candidate for the word ``futile''.

For the objective metrics, negated cosine distances between speaker embeddings (the O-SPK metrics) show negative or weak correlations with S-ACC-SMOS and S-SPK-SMOS. These results suggest that the embedding-based scores do not closely track perceived accent or speaker similarity in this evaluation, which uses content-matched reference--candidate pairs and enhanced, self-recorded reference speech. Recording conditions and speech enhancement may contribute to this mismatch, although their effects cannot be isolated from the present correlation analysis. O-ACC (Voxlect) shows the strongest objective--subjective association, moderately correlating with S-ACC-Freq ($\rho=0.46$). This finding suggests that Voxlect embeddings at the utterance level may be used as proxies for ranking utterances with different accent generation quality.

\vspace{-1em}
\subsection{Subword-Level Correlation}
\vspace{-1em}
\begin{table}[h!]
    \centering
    \setlength{\tabcolsep}{4pt}
    \renewcommand{\arraystretch}{1.1}
    \resizebox{0.75\columnwidth}{!}{
        \begin{tabular}{lccc}
        \toprule
        Subword-level Metric & Mean & Max & Midpoint \\
        \midrule
        O-F0 & \cellcolor{red!10}0.10 & \cellcolor{red!14}\textbf{0.14} & \cellcolor{red!8}0.08 \\
        O-Formants & \cellcolor{red!0}0.00 & \cellcolor{red!6}\textbf{0.06} & \cellcolor{red!2}0.02 \\
        O-PPG  & \cellcolor{red!7}0.07 & \cellcolor{red!15}\textbf{0.15} & \cellcolor{red!5}0.05 \\
        O-SPK (ECAPA-TDNN) & \cellcolor{red!8}0.08 & \cellcolor{red!15}\textbf{0.15} & \cellcolor{red!4}0.04 \\
        O-SPK (WavLM-SV) & \cellcolor{red!19}0.19 & \cellcolor{red!22}\textbf{0.22} & \cellcolor{red!16}0.16 \\
        O-ACC (GenAID) & \cellcolor{red!11}0.11 & \cellcolor{red!16}\textbf{0.16} & \cellcolor{red!9}0.09 \\
        O-ACC (Voxlect) & \cellcolor{red!14}0.14 & \cellcolor{red!22}\textbf{0.22} & \cellcolor{red!11}0.11 \\
        \bottomrule
        \end{tabular}
    }
    \vspace{-0.5em}
    \captionsetup{justification=centering}    
    \caption{\textit{Subword}-level \gls{srcc} between S-Anno-Score (proportion of listeners who select the segment) and objective distances across 4,140 subwords, using mean/max/midpoint scoring.}
    \label{tab:corr_seg}
\end{table}
\vspace{-1em}
\noindent Table~\ref{tab:corr_seg} shows that max aggregation gives the highest observed correlation for all seven metrics (Table~\ref{tab:corr_seg}). This pattern suggests that local peaks in objective dissimilarity may better reflect listeners' segmental perception. Nevertheless, correlations remain weak across all metrics and scoring methods, reaching at most $\rho=0.22$ for WavLM-SV and Voxlect. Thus, even with max aggregation, the evaluated objective metrics show limited alignment with perceived subword-level accent differences.

\vspace{-0.5em}
\section{Conclusions}
\vspace{-0.5em}

We presented a segmental accent evaluation protocol combining utterance-level similarity ratings with subword annotations from listeners in the target accent regions. Speaker and accent ratings remain strongly associated, while local annotations identify salient pronunciation differences and sometimes yield different system rankings. The seven evaluated objective distances show limited correspondence with these annotations, highlighting the challenges of relating existing representations to local perceptual judgements. These findings motivate evaluating accent generation through both utterance-level ratings and segmental annotations. Modest inter-annotator agreement and the limited number of speakers and samples evaluated under budget constraints warrant further validation. The released resources provide a basis for segmental annotation protocols and developing objective metrics better informed by listeners' perceptions.


\clearpage
\section{Acknowledgement}
This work was supported in part by the UKRI AI CDT in Responsible and Trustworthy in-the-world NLP (Grant EP/Y030656/1) and UKRI CDT in NLP (Grant EP/S022481/1), the University of Edinburgh.

\bibliographystyle{IEEEbib}
\bibliography{strings,refs}

@inproceedings{zhong2025pariwise,
  title     = {{Pairwise Evaluation of Accent Similarity in Speech Synthesis}},
  author    = {Jinzuomu Zhong and Suyuan Liu and Dan Wells and Korin Richmond},
  year      = {2025},
  booktitle = {{Proc. Interspeech}},
  pages     = {2290--2294},
  doi       = {10.21437/Interspeech.2025-2283},
  issn      = {2958-1796},
}

@inproceedings{zhong2025accentbox,
  author={Zhong, Jinzuomu and Richmond, Korin and Su, Zhiba and Sun, Siqi},
  booktitle={IEEE ICASSP}, 
  title={{AccentBox: Towards High-Fidelity Zero-Shot Accent Generation}}, 
  year={2025},
  volume={},
  number={},
  pages={1-5},
  doi={10.1109/ICASSP49660.2025.10888332}
}

@inproceedings{sanchez2026when,
    author = {Sanchez, Ariadna and Zhong, Jinzuomu and Deligianni, Artemis and Ross, Alice and King, Simon},
    title = {{When Text-to-Speech Speaks in Your Voice: A Study on Public Perception}},
    year = {2026},
    isbn = {9798400727412},
    address = {New York, NY, USA},
    url = {https://doi.org/10.1145/3816046.3816200},
    booktitle = {Proc. ACM CUI},
    articleno = {36},
    numpages = {17}
}

@inproceedings{gutierrez2021location,
  title     = {{Location, Location: Enhancing the Evaluation of Text-to-Speech synthesis using the Rapid Prosody Transcription Paradigm}},
  author    = {Elijah Gutierrez and Pilar Oplustil-Gallegos and Catherine Lai},
  year      = {2021},
  booktitle = {{11th Speech Synthesis Workshop}},
  pages     = {25--30},
  doi       = {10.21437/SSW.2021-5},
}

@inproceedings{minixhofer2026ttsds2,
  title={{TTSDS2: Resources and Benchmark for Evaluating Human-Quality Text to Speech Systems}},
  author={Minixhofer, Christoph and Klejch, Ondrej and Bell, Peter},
  booktitle={ICLR},
  volume={2026},
  pages={29228--29247},
  year={2026}
}

@inproceedings{wang2025maskgct,
    title={{MaskGCT: Zero-Shot Text-to-Speech with Masked Generative Codec Transformer}},
    author={Wang, Yuancheng and Zhan, Haoyue and Liu, Liwei and Zeng, Ruihong and Guo, Haotian and Zheng, Jiachen and Zhang, Qiang and Zhang, Xueyao and Zhang, Shunsi and Wu, Zhizheng},
    booktitle={Proc. ICLR},
    volume={2025},
    pages={47127--47150},
    year={2025}
}

@inproceedings{chen2025f5tts,
    title = "{{F5-TTS: A Fairytaler that Fakes Fluent and Faithful Speech with Flow Matching}}",
    author = "Chen, Yushen  and
      Niu, Zhikang  and
      Ma, Ziyang  and
      Deng, Keqi  and
      Wang, Chunhui  and
      Zhao, Jian  and
      Yu, Kai  and
      Chen, Xie",
    booktitle = "Proc. ACL",
    month = jul,
    year = "2025",
    address = "Vienna, Austria",
    url = "https://aclanthology.org/2025.acl-long.313/",
    doi = "10.18653/v1/2025.acl-long.313",
    pages = "6255--6271",
    ISBN = "979-8-89176-251-0",
}

@INPROCEEDINGS{li2025scalable,
    author={Xinyuan, Henry Li and Cai, Zexin and Garg, Ashi and Duh, Kevin and García-Perera, Leibny Paola and Khudanpur, Sanjeev and Andrews, Nicholas and Wiesner, Matthew},
    booktitle={IEEE ASRU}, 
    title={Scalable Controllable Accented TTS}, 
    year={2025},
    volume={},
    number={},
    pages={1-8},
    doi={10.1109/ASRU65441.2025.11434737}
}

@article{wang2026joycent,
  title={{Joycent: Diffusion-based Accent TTS without Accented Phone Prediction}},
  author={Wang, Xintong and Wang, Ye},
  journal={arXiv preprint arXiv:2606.16417},
  year={2026}
}

@inproceedings{michel2025s,
  title={{“It’s not a representation of me”: Examining Accent Bias and Digital Exclusion in Synthetic AI Voice Services}},
  author={Michel, Shira and Kaur, Sufi and Gillespie, Sarah Elizabeth and Gleason, Jeffrey and Wilson, Christo and Ghosh, Avijit},
  booktitle={Proc. ACM FAccT},
  pages={228--245},
  year={2025}
}

@book{wells1982accents,
  title={{Accents of English: Volume 1}},
  author={Wells, John C},
  volume={1},
  year={1982},
  publisher={Cambridge University Press}
}

@inproceedings{srinivasavaradhan2025the,
  title     = {{The State Of TTS: A Case Study with Human Fooling Rates}},
  author    = {Praveen {Srinivasa Varadhan} and Sherry Thomas and Sai {Teja M S} and Suvrat Bhooshan and Mitesh M. Khapra},
  year      = {2025},
  booktitle = {{Proc. Interspeech}},
  pages     = {2285--2289},
  doi       = {10.21437/Interspeech.2025-2765},
  issn      = {2958-1796},
}

@inproceedings{huang2026codecmosaccent,
  title     = {{CodecMOS-Accent: A MOS Benchmark of Resynthesized and TTS Speech from Neural Codecs Across English Accents}},
  author    = {Wen-Chin Huang and Nicholas Sanders and Erica Cooper},
  year      = {2026},
  booktitle = {{Proc. Interspeech}},
  pages     = {6130--6135},
  doi       = {10.21437/Interspeech.2026-1273},
  issn      = {2958-1796},
}

@inproceedings{schroeter2022deepfilternet2,
  title = {{DeepFilterNet2: Towards Real-Time Speech Enhancement on Embedded Devices for Full-Band Audio}},
  author = {Schröter, Hendrik and Escalante-B., Alberto N. and Rosenkranz, Tobias and Maier, Andreas},
  booktitle={17th International Workshop on Acoustic Signal Enhancement (IWAENC)},
  year = {2022},
}

@article{cho2024coding,
  title={{Coding Speech through Vocal Tract Kinematics}},
  author={Cho, Cheol Jun and Wu, Peter and Prabhune, Tejas S and Agarwal, Dhruv and Anumanchipalli, Gopala K},
  journal={IEEE Journal of Selected Topics in Signal Processing},
  volume={18},
  number={8},
  pages={1427--1440},
  year={2024},
  publisher={IEEE}
}

@article{chen2022wavlm,
	title = {{WavLM: Large-Scale Self-Supervised Pre-Training for Full Stack Speech Processing}},
	volume = {16},
	number = {6},
	journal = {IEEE Journal of Selected Topics in Signal Processing},
	author = {Chen, Sanyuan and Wang, Chengyi and Chen, Zhengyang and Wu, Yu and Liu, Shujie and Chen, Zhuo and Li, Jinyu and Kanda, Naoyuki and Yoshioka, Takuya and Xiao, Xiong and Wu, Jian and Zhou, Long and Ren, Shuo and Qian, Yanmin and Qian, Yao and Wu, Jian and Zeng, Michael and Yu, Xiangzhan and Wei, Furu},
	month = oct,
	year = {2022},
	pages = {1505--1518},
}

@inproceedings{desplanques2020ecapa,
  title     = {{ECAPA-TDNN: Emphasized Channel Attention, Propagation and Aggregation in TDNN Based Speaker Verification}},
  author    = {Brecht Desplanques and Jenthe Thienpondt and Kris Demuynck},
  year      = {2020},
  booktitle = {{Proc. Interspeech}},
  pages     = {3830--3834},
  doi       = {10.21437/Interspeech.2020-2650},
  issn      = {2958-1796},
}

@inproceedings{feng2026voxlect,
    author = {Feng, Tiantian and Huang, Kevin and Xu, Anfeng and Shi, Xuan and Lertpetchpun, Thanathai and Lee, Jihwan and Lee, Yoonjeong and Byrd, Dani and Narayanan, Shrikanth},
    title = {{Voxlect: A Speech Foundation Model Benchmark for Modeling Dialects and Regional Languages Around the Globe}},
    year = {2026},
    isbn = {9798400722585},
    address = {New York, NY, USA},
    url = {https://doi.org/10.1145/3770854.3785687},
    doi = {10.1145/3770854.3785687},
    booktitle = {Proc. ACM SIGKDD},
    pages = {2640–2651},
    numpages = {12},
    location = {Republic of Korea},
    series = {KDD '26}
}

@inproceedings{churchwell2024ppgs,
    author={Churchwell, Cameron and Morrison, Max and Pardo, Bryan},
    booktitle={IEEE ICASSP Workshops}, 
    title={{High-Fidelity Neural Phonetic Posteriorgrams}}, 
    year={2024},
    volume={},
    number={},
    pages={823-827},
    doi={10.1109/ICASSPW62465.2024.10669905}
}

@inproceedings{mcauliffe2017mfa,
  title     = {{Montreal Forced Aligner: Trainable Text-Speech Alignment Using Kaldi}},
  author    = {Michael McAuliffe and Michaela Socolof and Sarah Mihuc and Michael Wagner and Morgan Sonderegger},
  year      = {2017},
  booktitle = {Proc. Interspeech},
  pages     = {498--502},
  doi       = {10.21437/Interspeech.2017-1386},
  issn      = {2958-1796},
}

@inproceedings{richmond2009lts,
  title     = {{Robust LTS rules with the Combilex speech technology lexicon}},
  author    = {Korin Richmond and Robert A. J. Clark and Sue Fitt},
  year      = {2009},
  booktitle = {{Proc. Interspeech}},
  pages     = {1295--1298},
  doi       = {10.21437/Interspeech.2009-405},
  issn      = {2958-1796},
}

@inproceedings{ghorshi2006comparative,
  title     = {{Comparative Analysis of Formants of British, American and Australian Accents}},
  author    = {Seyed Ghorshi and Saeed Vaseghi and Qin Yan},
  year      = {2006},
  booktitle = {{Proc. Interspeech}},
  pages     = {paper 1252-Mon1BuP.9},
  doi       = {10.21437/Interspeech.2006-35},
  issn      = {2958-1796},
}

@article{deterding2006north,
  title={{The North Wind versus a Wolf: Short Texts for the Description and Measurement of English Pronunciation}},
  author={Deterding, David},
  journal={Journal of the International Phonetic Association},
  volume={36},
  number={2},
  pages={187--196},
  year={2006},
  publisher={Cambridge University Press}
}

@article{bishop2020phonology,
  title={Phonology, phonetics, and signal-extrinsic factors in the perception of prosodic prominence: Evidence from Rapid Prosody Transcription},
  author={Bishop, Jason and Kuo, Grace and Kim, Boram},
  journal={Journal of Phonetics},
  volume={82},
  pages={100977},
  year={2020},
  publisher={Elsevier}
}





\end{document}